\documentclass[journal=nalefd,manuscript=letter]{achemso}
\setkeys{acs}{usetitle=true} %this dark magic includes articles titles
\setkeys{acs}{keywords=true} %and keywords

\usepackage{amsmath,amssymb}
\usepackage{graphicx}
\usepackage{mciteplus}

\author{Ilya~Razdolski}
\affiliation{Fritz-Haber-Institut der MPG, Phys. Chemie, Faradayweg 4-6, 14195 Berlin, Germany}
\email{razdolski@fhi-berlin.mpg.de}
\author{Yiguo~Chen}
\affiliation{The Blackett Laboratory, Imperial College London, London SW7 2AZ, United Kingdom}
\altaffiliation{Department of Electrical and Computer Engineering, National University of Singapore, Singapore}
\author{Alexander~J.~Giles}
\affiliation{U.S. Naval Research Laboratory, Washington, DC, United States}
\author{Sandy~Gewinner}
\author{Wieland~Sch{\"o}llkopf}
\affiliation{Fritz-Haber-Institut der MPG, Phys. Chemie, Faradayweg 4-6, 14195 Berlin, Germany}
\author{Minghui~Hong}
\affiliation{Department of Electrical and Computer Engineering, National University of Singapore, Singapore}
\author{Martin~Wolf}
\affiliation{Fritz-Haber-Institut der MPG, Phys. Chemie, Faradayweg 4-6, 14195 Berlin, Germany}
\author{Vincenzo~Giannini}
\affiliation{The Blackett Laboratory, Imperial College London, London SW7 2AZ, United Kingdom}
\author{Joshua~D.~Caldwell}
\affiliation{U.S. Naval Research Laboratory, Washington, DC, United States}
\author{Stefan~A.~Maier}
\affiliation{The Blackett Laboratory, Imperial College London, London SW7 2AZ, United Kingdom}
\author{Alexander~Paarmann}
\affiliation{Fritz-Haber-Institut der MPG, Phys. Chemie, Faradayweg 4-6, 14195 Berlin, Germany}
\email{paarmann@fhi-berlin.mpg.de}

\title{Resonant enhancement of second harmonic generation in the mid-infrared using localized surface phonon polaritons in sub-diffractional nanostructures}
\begin{document}

\begin{abstract}
%Employing free electron laser radiation for the infrared SHG spectroscopy of SiC nanopillars, we report strong SHG enhancement associated with the excitation of the localised surface phonon-polaritons eigenmodes in the SiC Reststrahlen band.
We report on strong enhancement of mid-infrared second harmonic generation (SHG) from SiC nanopillars due to the resonant excitation of localized surface phonon-polaritons within the Reststrahlen band. 
%The measurements employ intense and tunable mid-infrared free-electron laser pulses for second harmonic generation  spectroscopy. 
The magnitude of the SHG peak at the monopole mode experiences a strong dependence on the resonant frequency beyond that described by the field localization degree and the dispersion of linear and nonlinear-optical SiC properties.
%We analyse the possible interplay of the localised surface phonon-polaritons with the zone-folded optic phonons in anisotropic SiC of different polytypes.
Comparing the results for the identical nanostructures made of 4H and 6H SiC polytypes, we demonstrate the interplay of localized surface phonon polaritons with zone-folded weak phonon modes of the anisotropic crystal. Tuning the monopole mode in and out of the region where the zone-folded phonon is excited in 6H-SiC, we observe a prominent increase of the already monopole-enhanced SHG output when the two modes are coupled. Envisioning this interplay as one of the showcase features of mid-infrared nonlinear nanophononics, we discuss its prospects for the effective engineering of nonlinear-optical materials with desired properties in the infrared spectral range.
\end{abstract}

Light localization in sub-wavelength volumes is a core of modern nanophotonics. Conventional methods of achieving strong confinement of the electromagnetic fields extensively utilize unique properties of surface plasmons. A remarkable variety of objects and materials supporting these excitations ensures the key role of plasmonics in a broad range of applications \cite{Barnes03,Maierbook07,Schuller10,Boriskina13,Brongersma15}. Apart from unparalleled sensitivity of plasmonic structures to the optical properties of the environment, strong light localization facilitates nonlinear-optical effects \cite{KauranenZayats,HentschelNLPlasm15,ButetACSNano15}. Owing to the spectral tunability of the localized plasmon resonances and their sizeable nonlinearity, metallic nanostructures of different shapes and sizes have earned their place in nonlinear photonics.

Despite obvious advantages of plasmon-based nanophotonics, metallic
nanoobjects exhibit significant optical losses, which lower the quality factor of the localized surface plasmon modes.
%\emph{and limit the operational frequencies to the UV--near-infrared (IR) spectral range}. 
%Although sensing applications based on linear optical effects have successfully overcome this drawback, t
%\emph{In the case of the former}, the efficiency of the nonlinear-optical processes is severely hampered, while 
Fast plasmon damping (typically on the order of 10 fs) due to ohmic losses \cite{VahalaNature03,BarnesJOPA06} thus inhibits nonlinear-optical conversion.
An alternative, promising metal-free approach has been suggested, utilizing polar dielectrics such as SiC \cite{GreffetNature02,HillenbrandNat02,NeunerOpEx07,CaldwellNanoLett13,YiguoACSPhot14} or BN \cite{CaldwellNatComm14,DaiScience14,CaldwellNatPhot15,LiNatComm15,DaiNatComm15} for high-quality light confinement in the mid-infrared (IR). In these materials, the subdiffractional confinement of electromagnetic radiation relies upon surface phonon polaritons (SPhP) in the Reststrahlen band: the electric polarization is created due to coherent oscillations of the ions instead of the electron or hole densities.
Due to the significantly longer scattering times associated with optical phonons as compared to surface plasmons, the lifetimes of SPhPs tend to be on the order of picoseconds, much longer than their plasmonic counterparts \cite{CaldwellReview14}.
% It has been shown that the quality of the SPhP excitations can exceed those of surface plasmons at least one order of magnitude
% 
In addition, due to energies associated with optical phonons, SPhPs with typical frequencies within the mid-IR ($>6\mu m$) to the THz domain hold high promise for spectroscopic and nanophotonic applications \cite{FergusonNatPhot02,UlbrichtRevModPhys11,KampfrathNatPhot13}.

In this Letter, we undertake a first step towards the largely unexplored domain of mid-IR nonlinear nanophononics. We study the
nonlinear-optical response of localized SPhPs using nanostructures made of different polytypes of SiC. Using free electron laser (FEL) radiation in the mid-IR spectral range \cite{Schollkopf15}, we probe second harmonic generation from rectangular arrays of sub-diffractional, cylindrical SiC nanopillars. The SHG yield in the Reststrahlen band of SiC demonstrates prominent enhancement at the wavelengths associated with the excitation of the SPhP eigenmodes of the pillars. Depending on both the size and the spatial periodicity of the pillars, the SHG-probed eigenmode exhibits a spectral shift accompanied with strong variations of the SHG intensity. Analyzing different SiC polytypes, we demonstrate the interplay of the localized SPhPs with the zone-folded optical phonon modes. We further conclude that strong coupling of the two modes allows for a significant additional  modulation of the SPhP-enhanced SHG output.
%conclude that the interplay of the SPhP eigenmodes with the hidden weak optical phonon modes \cite{x} greatly enhances the SHG output.

\begin{figure}[t]
  \includegraphics[width=0.95\textwidth]{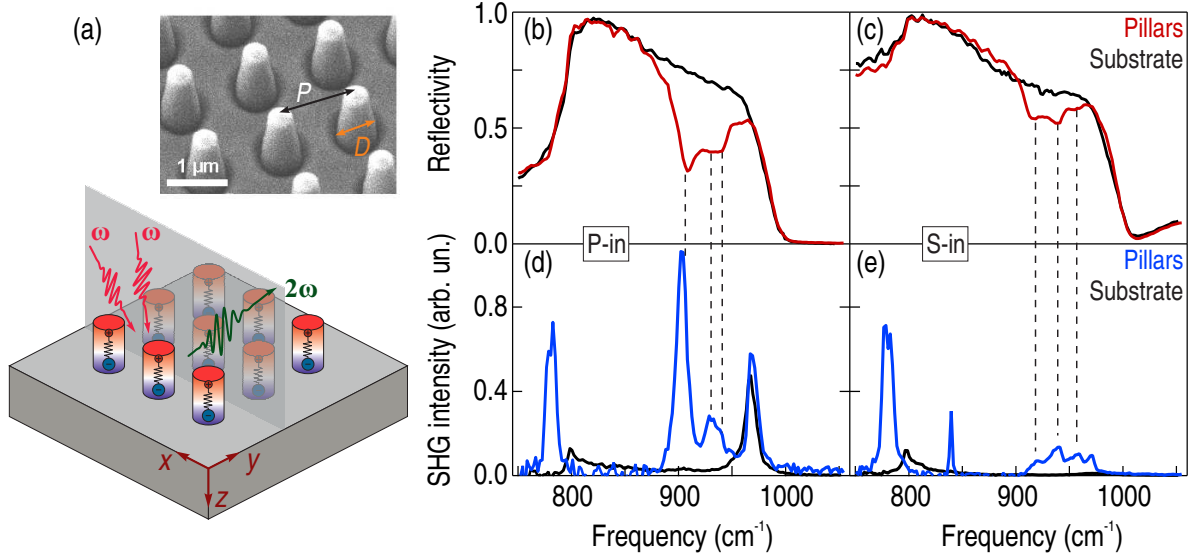} 
	\caption{
	(a) Schematic of the experimental approach. The inset shows an electron microscopy image of the nanopillar array.
	(b-c) Linear reflectivity spectra for \emph{p}- and \emph{s}-polarised incident radiation obtained on the array of 4H-SiC nanopillars (red) and on the 4H-SiC substrate (black).
	(d-e) SHG output spectra on the 4H-SiC nanopillars (blue) and the 4H-SiC substrate (black). The vertical dashed lines indicate the excited eigenmodes of the pillars.
	}
	\label{general}
\end{figure}

The schematic of our experimental approach is outlined in Fig.~\ref{general},a. We employed a non-collinear SHG configuration discussed elsewhere \cite{PaarmannAPL15} to perform spectroscopic SHG measurements on square arrays of 1~$\mu$m-tall 4H-SiC and 6H-SiC pillars with the main axis of the arrays in the \emph{xz} plane of incidence. The fundamental radiation incident at 28 and 62 degrees with respect to the normal to the sample surface was focused onto the sample with a peak fluence on the order of 10 mJ/cm$^2$. Both 4H and 6H-SiC samples were c-cut so that the \emph{c}-axis of the crystals was parallel to the surface normal.
Typical SHG and linear reflectivity spectra collected using the FEL radiation for the two incident polarisations are presented in Fig.~\ref{general},b-e. There, the respective spectra of the bare substrate are shown for comparison. For {\it p}-polarised fundamental radiation (Fig.~\ref{general},d),
% the SHG response features two pronounced peaks corresponding to the excitation of transverse and longitudinal optic phonons in SiC at the edges of the Reststrahl band 
the SHG response features two pronounced peaks located at the zone-center frequencies of transverse and longitudinal optical phonons in SiC \cite{PaarmannAPL15}, around 797 cm$^{-1}$ and 965 cm$^{-1}$, respectively. The corresponding SHG spectrum from the nanopillars demonstrate a much stronger SHG signal at around 900 cm$^{-1}$. Due to the absence of this peak in the SHG spectrum when the fundamental radiation is \emph{s}-polarised
%and the results of numerical simulations (see below)
, we attribute this SHG feature to the excitation of the monopole SPhP mode in the nanopillars \cite{YiguoACSPhot14}.

In general, the outgoing SHG field $\vec{E}^{2\omega}$ is related to the incident electromagnetic fields $E_i^{\omega}$ via the so-called local field factors $L_i^{\omega}$:

\begin{equation}
E_i^{2\omega}\propto \mathcal{P}_i^{2\omega}=\chi^{(2)}_{ijk}:(L_j^{\omega}E_j^{\omega})(L_k^{\omega}E_k^{\omega}),
\label{localfields}
\end{equation}
where $\mathcal{P}_i^{2\omega}$ is the nonlinear polarisation and $\chi^{(2)}_{ijk}$ is the nonlinear susceptibility tensor. The excitation of the SPhP monopole mode leads to strong localization of the \emph{z}-projection of the fundamental electric field $E_z$ (normal to the surface plane) and thus a resonant enhancement of $L_z^{\omega}$. The latter results in a pronounced increase of the SHG output when the fundamental radiation is \emph{p}-polarised.
% and explains why the SPhP monopole mode is not excited with the \emph{s}-polarised light. 
%Unlike the SPhP monopole mode which is not excited with the \emph{s}-polarised light.
However, the SPhP dipole modes observed in the range of 920--960~cm$^{-1}$ rely on the resonant enhancement of the in-plane electric fields described by the local field factors $L^{\omega}_{x,y}$ and thus can be excited with both \emph{p}- and \emph{s}-polarised fundamental radiation.
The total SHG response is given by a vector sum of the terms on the right hand side of Eq.~(1)
%nonlinear polarisation $P(2\omega)$ 
originating from various tensor components of the nonlinear susceptibility $\chi^{(2)}$. As the strength of $\chi^{(2)}_{zzz}$ is the largest in this spectral range \cite{PaarmannArXiv16}, the SPhP monopole mode, with an enhancement of $L_z^{\omega}$, naturally results in a higher SHG output than the dipole modes.

\begin{figure}[h]
  \includegraphics[width=0.47\textwidth]{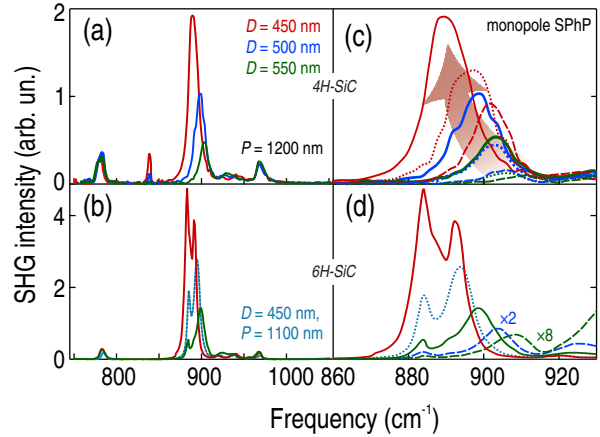} 
	\caption{
	Experimental \emph{p}-in SHG spectra for three different values of the pillar diameter \emph{D} for the 4H-SiC (a) and 6H-SiC (b) samples.
	(c-d) Zoom-in into the SHG spectra in the vicinity of the SPhP monopole resonance. 
	%The data for three different array periodicities \emph{P} is shown (c). 
	Here, the colours are the same as in (a,b), and solid, dotted, and dashed lines represent $P$ = 1200, 1100 and 1000 nm, respectively. 
	%The parameters of the 6H-SiC samples shown in (d) are the following: $D=0.45 \mu m$, $P=1.2\mu m$; $D=0.55 \mu m$, $P=1.2\mu m$; $D=0.45 \mu m$, $P=1.1\mu m$; $D=0.55 \mu m$, $P=1.0\mu m$; $D=0.5 \mu m$, $P=1.0\mu m$.
	}
	\label{shifts}
\end{figure}

\begin{figure*}[t]
  \includegraphics[width=0.95\textwidth]{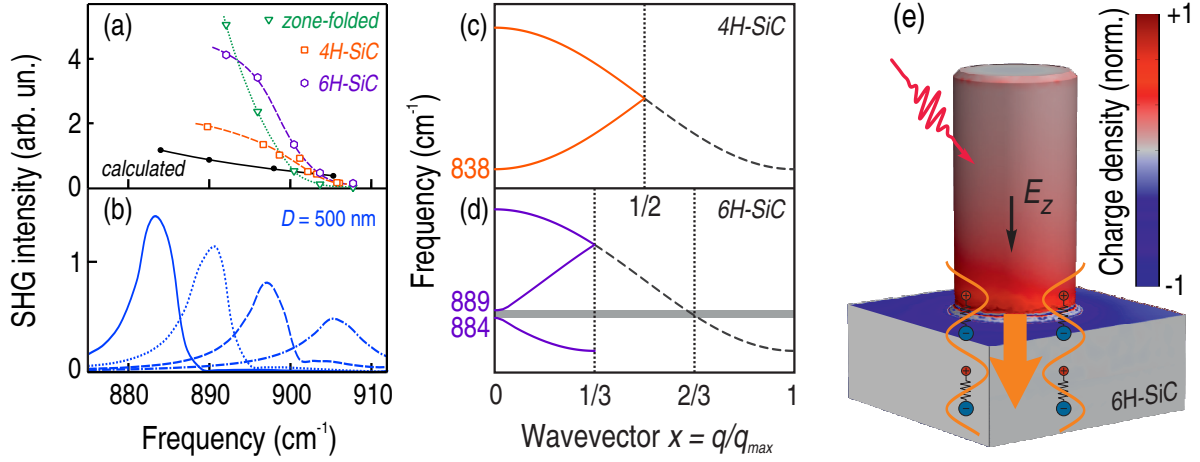} 
	\caption{
	(a) Variations of the resonant SHG output with the fundamental frequency of the SPhP monopole mode. The data for both 4H-SiC (red squares) and 6H-SiC (blue circles)demonstrate enhancement of the SPhP monopole resonance-driven SHG when the frequency of the SPhP mode is red-shifted. The full black circles indicate the expected SHG output modulation, as obtained from the numerical simulations (b). The green triangles demonstrate the increase of the SHG output driven by the excitation of the zone-folded mode in 6H-SiC samples.
	(b) Numerically simulated SHG output spectra in the vicinity of the SPhP monopole mode for the pillar diameter \emph{D}=500 nm and four different array periodicities \emph{P} = 900, 1000, 1100 and 1200 nm.
	(c-d) Sketch of the dispersion of the longitudinal optical phonon modes in the full Brillouin zone and the emergence of the zone-folded modes in 4H- and 6H-SiC polytypes.
	(e) Illustration of the coupling mechanism of the zone-folded and the monopole SPhP modes via the enhanced normal to surface projection of the electric field $E_z$. The false colour map in the background represents the calculated distribution of the electric charge when the SPhP monopole mode is excited.
	}
	\label{analysis}
\end{figure*}

The results of the systematic studies of the SHG response of various arrays of nanopillars are summarized in Fig.~\ref{shifts} 
%Here we discuss the experimental SHG spectra obtained
for the 4H-SiC (a,c) and 6H-SiC (b,d) samples.
The panels (a-b) illustrate the evolution of the SHG spectra upon varying pillar diameter $D$. It is seen that upon decreasing $D$, the SPhP monopole-driven SHG peak exhibits a clear redshift. Remarkably,  while the SPhP monopole mode shifts in the range of about 890-910 cm$^{-1}$, the SHG enhancement factor associated with the excitation of the SPhP monopole mode varies strongly with $D$. The panels (c-d) offer a zoom-in into the evolution of the SPhP monopole mode-driven SHG for a large variety of nanopillar arrays, indicating that the variations of the SHG enhancement are observed for both 4H and 6H samples.  The dependences of the SPhP monopole-driven SHG output on the spectral position of the SPhP monopole mode for the two SiC polytypes are shown in Fig.~\ref{analysis},a with open symbols. 

%It is seen that the SHG enhancement factor associated with the excitation of the SPhP monopole mode varies strongly with $D$ while the SPhP monopole mode itself shifts in the range of about 890-900 cm$^{-1}$. The panels (c-d) offer a zoom-in into the evolution of the SPhP monopole mode-driven SHG for a large variety of nanopillar arrays. 
%здесь обсуждение рис. 2, усиление ВГ, сдвиг
%It is seen that upon decreasing $D$, the SPhP monopole-driven SHG peak exhibits a clear redshift (Fig.~\ref{shifts},c-d). Remarkably, this spectral shift of the SPhP monopole mode is accompanied by a strong increase of the resonant SHG output. 
%As such, the spectral tunability of the eigenmode frequency with changing geometric parameters results in a strong modulation of the SHG conversion efficiency (Fig.~\ref{analysis},d).

%теория (кратко)

%It is also seen that it is not $P$ or $D$ on their own that determine the SHG enhancement. On the contrary, the role of these parameters is limited to the fact that they both determine the resonant frequency of the SPhP monopole mode. Once the frequency is set, for any combination of $P$ and $D$ the SHG output is found to be nominally independent of their particular values (see the dashed line in Fig.~\ref{analysis},d).

The strong dispersion in the SiC Reststrahlen band suggests that the observed spectral dependence of the SHG enhancement could be captured in numerical simulations. 
%%%%%%%%%%
%To obtain the SHG output, we adopted 
%
We calculated SHG response using both linear and nonlinear SiC dispersion\cite{CaldwellNanoLett13,PaarmannAPL15,PaarmannArXiv16}, the results of the simulation of the linear optical response\cite{YiguoACSPhot14}, and nonlinear polarization $\mathcal{P}^{2\omega}$ from Eq.~(1) spatially integrated over the SiC volume, with the following non-zero components of the nonlinear susceptibility: $\chi^{(2)}_{zzz}$, $\chi^{(2)}_{zxx}$ and $\chi^{(2)}_{xxz}=\chi^{(2)}_{xzx}$. The resultant SHG spectra simulated with COMSOL multiphysics software (www.comsol.com) are shown in Fig.~\ref{analysis},b with full symbols.
It is seen that the steep experimental dependence cannot be quantitatively described within the simple model used in the calculations, which yields only a moderate increase of the SHG output when the SPhP monopole mode frequency is decreased.

We note the work of Carpetti \emph{et al}. \cite{CarpettiOpEx12} where the authors thoroughly examined plasmon-induced SHG enhancement from arrays of Au nanoparticles as a function of the inter-particle distance $b$. In the regime of $b<\lambda$ (as it is in our case), the dependence of the SHG output on $b$ was explained in terms of the changing filling factor (and thus the associated number of active nanoemitters).  For very small inter-particle gaps ($b/\lambda\sim 10^{-2}$), a modulation of the SHG output has been  attributed to the modification of the electromagnetic field localization in the gaps \cite{CanfieldNanoLett07}. Further, a large mismatch between the spatial period of the nanopillars $P\sim 1 \mu m$ and the resonant light wavelength $\lambda\sim 10 \mu m$ rules out the excitation of propagating surface polaritons \cite{raether} which are known to enhance the SHG output \cite{AgarwalPRB82,CoutazJOSAB87,QuailJOSAB88,RazdolskiACS15,GubbinPRL16}.
Since all these effects are included in our simulations, we conclude that the origin of the observed SHG enhancement is unrelated to the periodicity of the array.

An alternative scenario for the observed trend could invoke modifications of the non-local SHG contribution enhanced by a regular array structure.
The importance of the non-local SHG was already demonstrated in a number of subwavelength-separated plasmonic nanoobjects \cite{ShanPRB06,CanfieldNanoLett07,KujalaPRL07,BachelierPRB10,ButetPRL10,RazdolskiPRB13,KrukACSPhot15,HilleJPC16}. The amplitude and phase of this non-local SHG source depends on the electric field distribution, i.e. on both the pillar size $D$ and the periodicity $P$. As such, the interference conditions between the SHG sources vary for different samples, thus resulting in a strong modification of the SHG output.

%сравнение 4H vs 6H 
Further, the apparent difference in the shape of the resonant SHG output for the 4H and 6H samples (see Fig.~\ref{shifts}) is related to the anisotropic nature of SiC. The hexagonal SiC polytypes are known to exhibit zone-folded weak modes in the Reststrahlen region \cite{BluetMSE99} originating from particular stacking of the atomic layers along the \emph{c}-axis of the crystal \cite{PatrickPR67,FeldmanPR68b}. These weakly IR-active zone-folded modes can be visualised in the reflectivity measurements at oblique incidence \cite{EngelbrechtPRB93}.  
%In a striking agreement to this picture, in those arrays of nanopillars where the SPhP monopole resonance approached the frequencies of the zone-folded modes, a complex double-peak structure of the resonant SHG response accompanied with its strong enhancement was found (Fig.~\ref{shifts},d).
%
Although zone-folded modes exist in both 4H and 6H polytypes, different stacking of the SiC atomic layers 
%results in  a significant variation of the properties of the former,
is responsible for them having different frequencies,
%This difference in the frequencies of the zone-folded modes in 4H-SiC and 6H-SiC polytypes is
as illustrated in Fig.~\ref{analysis},c-d. The additional periodicity in the crystal results in folding of the large Brillouin zone thus modifying the phonon dispersion and making the excitation of phonons with non-zero wavevectors ($q\neq 0$) possible. 
%For example, in the case of 6H-SiC, the zone-folded mode with the reduced wavevector $q=2/3$ $\pi/a$ can be optically excited.
%is split into two, where $a$ is the inter-atomic distance.
%which can be visualised in the reflectivity measurements at oblique incidence \cite{EngelbrechtPRB93}.
It is seen in Fig.~\ref{analysis},d that the zone-folded mode in the 6H-SiC polytype with the reduced wavevector $q=2/3$ $\pi/a$ can be excited 
%at about $884$ and $889$~cm$^{-1}$
in the range of $880-890$~cm$^{-1}$ which is close to the typical monopole SPhP resonant frequency of the SiC nanopillars discussed above.  In particular, the interaction of the SPhP and zone-folded mode which shifted the apparent spectral positions of the monopole SPhP eigenmode in the linear response \cite{CaldwellNanoLett13}, is seen responsible for the complex structure of the resonant SHG output in our experiments (Fig.~\ref{shifts},d).
The weak IR activity of the zone-folded modes is related to the large negative dielectric permittivity of SiC in the Reststrahlen band. As such, the out-of-plane component of the electromagnetic field $E_z$ remains small which inhibits the coupling of incident light to the zone-folded phonon mode. However, the excitation of the SPhP monopole mode in the nanopillars drives a strong increase of $E_z$, which facilitates the SPhP monopole interaction with the zone-folded phonon (Fig.~\ref{analysis},c).

Figure~\ref{analysis},a illustrates the effect of the zone-folded mode on the observed SHG output. The open red squares (4H) and blue circles (6H) depict the SHG intensity obtained at the SPhP monopole (resonant) frequencies, and the dashed lines illustrate a clear correlation between the maximum SHG output and the spectral position of the SPhP monopole peak. Similar trends are observed for the 4H and 6H-SiC samples as long as the SPhP monopole and the zone-folded mode are well-separated. When these two resonances start to overlap, an additional enhancement of the SHG output produced at the SPhP monopole resonance in 6H-SiC samples is observed. Moreover, the dependence of the  SHG signal at the frequency of the zone-folded mode $884$~cm$^{-1}$ (green triangles) exhibits a much faster increase when the two resonances are brought together (dotted line), indicating an efficient interplay between the SPhP monopole and the zone-folded mode.

We note that the interaction of the localized SPhP eigenmodes and the intrinsic excitations of the medium is a unique fingerprint of mid-IR nanophononics. Indeed, surface plasmon excitations in metals rely on the free electron gas, which is essentially isotropic. 
%Its weak interaction with the anisotropic crystal lattice leads to the increase of plasmon losses and usually hampers the SHG output. 
As such, the SHG output of plasmonic nanostructures is (i) largely determined by the metal of choice, usually Au, (ii) exhibits only weak spectral dependence \cite{CarpettiOpEx12,MetzgerNanoLett15} and (iii) is limited by robust phase relations in the likely case of multiple SHG sources \cite{HusuMeta08,PalombaPRL08,CarpettiPlasm14,KrukACSPhot15,RazdolskiACSPhot16}. On the contrary, the flexibility of the SPhPs is provided by the coupling of the surface phonon polariton excitations to the intrinsic bulk phonon modes. 
The latter can be engineered by designing artificial metamaterials based on hybrid multilayer structures \cite{DaiNano15,CaldwellNano16}, thus allowing for an effective control of their optical properties.
%Remarkably, the SHG output associated with the excitation of the zone-folded mode at $\approx 883$~cm$^{-1}$ experienced a similar enhancement. Despite the apparent double-peak shape of the SHG spectra, we were unable to fit the experimental data with the simple sum of the Lorentzian-shaped peaks. This strongly suggests that the interaction of the zone-folded mode with the SPhP resonance is crucial for understanding the prominent SHG enhancement discussed above. This interaction is strongly facilitated by the distribution of the electric field. The SPhP monopole resonance leads to the strong enhancement of the normal component of the electric field $E_z$, while the axial optic phonon constituting the zone-folded mode is characterized by the electric field along the \emph{c}-axis of the crystal. In the \emph{c}-cut SiC samples used in our work these directions coincide (Fig.~\ref{analysis},f), thus promoting the coupling of these modes and enhancement of the nonlinear-optical effects. 

To summarize, we have observed SHG output enhancement associated with the excitation of the SPhP eigenmodes in an array of nanopillars grown from SiC of different polytypes. The strongest SHG output is associated with the excitation of the monopole SPhP mode characterized by strong localization of the normal to surface projection of the electric field $E_z$. The spectral positions of the SHG peaks shift according to the geometric parameters of the nanophononic structures. We found a strong dependence of the magnitude of the SHG enhancement on the resonant frequency. This experimentally observed dependence cannot be quantitatively described by simply taking into account the sub-diffractional field localization and the dispersion of the linear and nonlinear optical properties of SiC.
Further, we discuss the interplay of the SPhPs and the intrinsic crystalline anisotropy for an efficient nonlinear-optical conversion. This mechanism is supported by the SHG spectral measurements on the 6H-SiC nanopillars where excitation of the SPhP monopole mode interacting with the weak zone-folded phonon resulted in an additional enhancement of the SHG output. 
The presence of intrinsic resonances strongly alters the phase relations in their vicinity, providing a natural way for optimizing the SHG response in a relatively narrow spectral range. 
Our findings demonstrate high potential of mid-IR nonlinear nanophononics as a novel and promising platform for nonlinear optics and illustrate the rich opportunities it provides for efficient control over nonlinear-optical response.

\begin{acknowledgement}
The authors thank G.~Kichin and A.~Kirilyuk (Radboud University Nijmegen) for stimulating discussions. S.A.M. acknowledges the Office of Naval Research, the Royal Society, and the Lee-Lucas Chair in Physics. A.J.G. acknowledges financial support from the NRC/NRL Postdoctoral Fellowship Program. Funding for J.D.C. was provided by the Office of Naval Research through the Naval Research Laboratory's Nanoscience Institute.                    
\end{acknowledgement}

%\suppinfo

%{\it Supporting Information Available}:
%The Supporting Information is available free of charge on the ACS Publications website at DOI: 10.1021/acsphotonics.XXXXXXX.

%\title{Enhancement of second harmonic generation from sub-wavelength localised surface phonon-polaritons in SiC nanopillars}
\section{Supplementary Information:\\
Numerical simulations}
%Resonant enhancement of second harmonic generation in the mid-infrared using localized surface phonon polaritons in sub-diffractional nanostructures}

%\section{}

The simulations we discuss here allowed us to obtain spectra of the SHG response. We employed COMSOL multiphysics software (www.comsol.com) in order to perform simulations of the linear optical response as described in Ref.~\cite{YiguoACSPhot14}. A unit cell containing a single SiC pillar attached to the SiC substrate was constructed, with Floquet boundary conditions along the $x$-axis and periodic boundary conditions along the $y$-axis, perpendicular to the plane of incidence. 
A \emph{p}-polarized plane wave at frequency $\omega$ was launched towards the SiC structure at an incident angle of 25$^{\circ}$.
Originating from the \emph{p}-polarised light source, the electric field $\vec{E}$ inside the SiC pillar and substrate was recorded.
Then, this electric field was translated into the nonlinear polarization $\mathcal{P}^{2\omega}$, according to Eq.~(1) in the main Manuscript. There, the following non-zero components of the nonlinear susceptibility were accounted for: $\chi^{(2)}_{zzz}$, $\chi^{(2)}_{zxx}$ and $\chi^{(2)}_{xxz}=\chi^{(2)}_{xzx}$. 

As a next step, $\mathcal{P}^{2\omega}$ in the pillar and the substrate was regarded as the source of scattered electric field $E^{2\omega}$ inside the unit cell. The SHG output was then obtained by integrating the power flow through a \emph{xy}-plane set above the substrate. Sweeping the fundamental frequency and taking the dispersion of both linear and nonlinear SiC susceptibilities from Refs.\cite{CaldwellNanoLett13,PaarmannAPL15,PaarmannArXiv16}, we calculated several SHG spectra for samples with various periodicity of the pillars. We note that in these calculations, the input power density at fundamental frequency was kept constant, and the SHG output power was normalized to the area of the \emph{xy}-plane $S\propto P^2$. The results of the simulations presented in Fig.~\ref{analysis} in the main Manuscript demonstrate a single pronounced peak corresponding to the excitation of the SPhP monopole mode.

%\bibliography{Nanopillars}

\end{document}